\documentstyle[12pt]{article}

\input epsf

\newlength{\extralineskip}

\newcommand{\beq}{\begin{equation}}
\newcommand{\eeq}{\end{equation}}
\newcommand{\bd}{\begin{displaymath}}
\newcommand{\ed}{\end{displaymath}}
\def\bea{\begin{eqnarray}}
\def\eea{\end{eqnarray}}

\def\ba{\beq\new\begin{array}{c}}
\def\ea{\end{array}\eeq}

\def\sympl{{\cal M}}
\def\lie{{\cal L}}
\def\pfaff{{\rm Pfaff}}
\def\hess{{\cal H}}
\def\inbar{\,\vrule height1.5ex width.4pt depth0pt}
\def\IC{\relax\hbox{$\inbar\kern-.3em{\rm C}$}}
\def\IR{\relax{\rm I\kern-.18em R}}
\def\IZ{{{\rm Z}\!\!{\rm Z}}}

\def\tr{{\rm tr}}
\def\e{~{\rm e}}
\makeatletter
\newdimen\normalarrayskip              
\newdimen\minarrayskip                 
\normalarrayskip\baselineskip
\minarrayskip\jot
\newif\ifold             \oldtrue            \def\new{\oldfalse}
\def\arraymode{\ifold\relax\else\displaystyle\fi} 
\def\@arrayskip{\ifold\baselineskip\z@\lineskip\z@
     \else
     \baselineskip\minarrayskip\lineskip2\minarrayskip\fi}
\def\@arrayclassz{\ifcase \@lastchclass \@acolampacol \or
\@ampacol \or \or \or \@addamp \or
   \@acolampacol \or \@firstampfalse \@acol \fi
\edef\@preamble{\@preamble
  \ifcase \@chnum
     \hfil$\relax\arraymode\@sharp$\hfil
     \or $\relax\arraymode\@sharp$\hfil
     \or \hfil$\relax\arraymode\@sharp$\fi}}
\def\@array[#1]#2{\setbox\@arstrutbox=\hbox{\vrule
     height\arraystretch \ht\strutbox
     depth\arraystretch \dp\strutbox
     width\z@}\@mkpream{#2}\edef\@preamble{\halign \noexpand\@halignto
\bgroup \tabskip\z@ \@arstrut \@preamble \tabskip\z@ \cr}%
\let\@startpbox\@@startpbox \let\@endpbox\@@endpbox
  \if #1t\vtop \else \if#1b\vbox \else \vcenter \fi\fi
  \bgroup \let\par\relax
  \let\@sharp##\let\protect\relax
  \@arrayskip\@preamble}

\begin{document}
\begin{titlepage}
\setcounter{footnote}0
\rightline{\baselineskip=12pt\vbox{\halign{&#\hfil\cr
&UBC/S-95/1\cr
&OUTP-95-46P\cr
&hep-th/9511199 &\cr
{   }&\cr &\today\cr}}}
\vspace{0.5in}
\begin{center}
{\Large\bf Conformal Motions and the Duistermaat-Heckman Integration
Formula\footnote{\baselineskip=12pt This work was supported in part by the
Natural Sciences and Engineering Research Council of Canada.}}\\
\medskip
\vskip0.5in
\baselineskip=12pt

\normalsize {\bf Lori D. Paniak} and {\bf Gordon W. Semenoff}
\medskip

\baselineskip=12pt

{\it Department of Physics, University of British Columbia\\
Vancouver, British Columbia, Canada V6T 1Z1}

\bigskip
\medskip

{\bf Richard J. Szabo}
\medskip

\baselineskip=12pt

{\it Department of Theoretical Physics, University of Oxford\\ 1 Keble Road,
 Oxford OX1 3NP, U.K.}

\end{center}
\vskip1.5in

\begin{abstract}
\baselineskip=12pt

We derive a geometric integration formula for the partition function of a
 classical dynamical system and use it to show that corrections to the WKB
 approximation vanish for any Hamiltonian which generates conformal motions of
 some Riemannian geometry on the phase space. This generalizes previous cases
where the Hamiltonian was taken as an isometry generator. We show that this
conformal symmetry is similar to the usual formulations of the
 Duistermaat-Heckman integration formula in terms of a supersymmetric Ward
 identity for the dynamical system. We present an explicit example of
a localizable Hamiltonian system in this context and use it to demonstrate how
 the dynamics of such systems differ from previous examples of the
 Duistermaat-Heckman theorem.

\end{abstract}
\end{titlepage}
\newpage

\setcounter{footnote}0
\baselineskip=18pt

Understanding the circumstances under which a partition function
is given exactly by its semi-classical approximation reveals connections
between classical mechanics and the geometry and topology of the
 associated phase space. The Duistermaat-Heckman theorem \cite{DH}--\cite{bv}
has been the basis of so-called localization theory in both mathematics and
physics \cite{atiyah}--\cite{szsem2} (see \cite{blau-thom} for a recent
review) and it provides geometric criteria for the exactness of the
semi-classical approximation for the finite-dimensional partition function
\beq
Z(T)=\int_\sympl d^{2n}x~\sqrt{\det\omega(x)}\e^{iTH(x)}
\label{clpart}\eeq
which describes the statistical dynamics (with imaginary temperature) of a
classical Hamiltonian system. Here $\sympl$ is a $2n$-dimensional phase space
and $\omega_{\mu\nu}(x)$ is an antisymmetric tensor field on $\sympl$ which is
non-degenerate, $\det\omega(x)\neq0~~\forall x\in\sympl$, and whose matrix
inverse $\omega^{\mu\nu}$ defines the Poisson brackets
$\{x^\mu,x^\nu\}=\omega^{\mu\nu}(x)$ of the dynamical system. $H$ is a smooth
Hamiltonian function on $\sympl$ which for simplicity we assume has a finite
set of critical points $I(H)=\{p\in\sympl:dH(p)=0\}$ each of which is
non-degenerate.

The partition function (\ref{clpart}) has an
asymptotic expansion for large-$T$ with coefficients determined by the method
of stationary-phase
 approximation \cite{hormander}. The Duistermaat-Heckman theorem states that if
the phase space $\sympl$ is compact and closed and the classical time-evolution
$x(t)$ of the dynamical system traces out a torus $(S^1)^m$ in $\sympl$, then
the partition function (\ref{clpart}) is given {\it exactly} by the leading
order term
\beq
Z^{(0)}(T)=\left(\frac{2\pi}{T}\right)^n\sum_{p\in
I(H)}\e^{i\frac{\pi}{4}\eta_H(p)}
\sqrt{\frac{\det\omega(p)}{\det\hess(p)}}\e^{iTH(p)}
\label{dhthm}\eeq
of its stationary-phase loop-expansion \cite{DH}. Here
 $\hess(x)_{\mu\nu}\equiv\partial_\mu\partial_\nu H(x)$ is the non-degenerate
 Hessian matrix of $H$ and $\eta_H(p)$ is the spectral asymmetry of the Hessian
 at $p$ (the difference between the number of positive and negative eigenvalues
 of $\hess(p)$).

A fundamental assumption that leads to the Duistermaat-Heckman theorem
 and its generalizations \cite{bv} is the
existence of a  globally-defined metric tensor
$g=\frac{1}{2}g_{\mu\nu}(x)dx^\mu dx^\nu$ on $\sympl$ which is invariant under
the classical flows $x(t)\in\sympl$ of the Hamiltonian system, i.e.
\beq
g(x(t))=g(x(0))
\label{ginv}\eeq
The condition (\ref{ginv}) is a very restrictive one on the Hamiltonian
dynamics as it implies
 that $H$ must generate a global $U(1)$-action on $\sympl$ \cite{niemi-tirk}.
 The set of Hamiltonian systems which obey these constraints has been examined
in \cite{dlr,szsem1,szsem2}. In this Letter we will show that the
Duistermaat-Heckman integration formula
(\ref{dhthm}) for the partition function (\ref{clpart}) still holds when the
geometric assumption (\ref{ginv}) is replaced by the weaker
 condition that there exists a metric tensor $g$ on $\sympl$ which is
 invariant under the classical time evolution of the dynamical system up to a
 change of scale,
\beq
g(x(t))=\e^{t\Lambda(x(t))}~g(x(0))
\label{gconfinv}\eeq
for some smooth function $\Lambda(x)$ on $\sympl$.
  We shall argue that this extended geometric requirement is similar
 to the isometry condition (\ref{ginv}) from the point of view of localizing
 (\ref{clpart}) onto the critical point set $I(H)$, except that the classical
dynamics now possess
 behaviour not normally observed for systems whose partition functions
 can be localized. We illustrate these features for some explicit examples
which show how the extension (\ref{gconfinv}) expands the set of previously
 studied Hamiltonian systems for which the Duistermaat-Heckman formula holds.

First, we will discuss some features of the integration in (\ref{clpart}). We
describe the exterior differential calculus of the manifold $\sympl$ by
introducing a set of
 anticommuting Grassmann variables $\psi^\mu$ which are to be identified
locally
 with the basis elements $\psi^\mu\sim dx^\mu$ of the cotangent bundle
 $T^*\sympl$ of $\sympl$. A differential $m$-form is represented by contracting
a rank-$m$ antisymmetric tensor function on $\sympl$ with
$\psi^{\mu_1}\cdots\psi^{\mu_m}$ and it can be regarded as a function
$\eta(x,\psi)=\frac{1}{m!}\eta_{\mu_1\cdots\mu_m}(x)
\psi^{\mu_1}\cdots\psi^{\mu_m}$
 on the super-manifold $\sympl\otimes T^*\sympl$.
 The integration of differential forms is defined on $\sympl\otimes
 T^*\sympl$ by introducing the usual Berezin rules for integrating Grassmann
 variables, $\int d\psi^\mu~\psi^\mu=1,\int d\psi^\mu~1=0$. With these rules,
we
 can absorb the determinant of the symplectic 2-form
 $\omega\equiv\frac{1}{2}\omega_{\mu\nu}(x)\psi^\mu\psi^\nu$ into the
 exponential in (\ref{clpart}) and write the classical partition function as
\beq
Z(T)=\frac{1}{(iT)^n}\int_{\sympl\otimes T^*\sympl}d^{2n}x~
d^{2n}\psi~\e^{iTS(x,\psi)}\equiv\int_\sympl\alpha
\label{clpartpsi}\eeq
where we have introduced the inhomogeneous differential form
\beq
\alpha =\frac{1}{(iT)^n} \e^{ i T ( H + \omega)} =
\e^{iTH}\sum_{k=0}^{n}\frac{(iT)^{k-n}}{k!}\omega^k
\label{alpha}\eeq
{}From the Berezin rules the integration in (\ref{clpartpsi}) is non-zero only
on
the top-form (degree $2n$) component $\omega^{n}$ of $\alpha$, and all forms in
(\ref{alpha}) of Grassmann-degree higher than $2n$ vanish because of the
fermionic nature of the variables $\psi^\mu$. The integral in (\ref{clpartpsi})
can be thought of as the partition function of a zero-dimensional quantum field
theory with bosonic fields $x^\mu$, fermion fields $\psi^\mu$, and action
$S(x,\psi)\equiv H(x)+\omega(x,\psi)$.

The Hamiltonian vector field is defined by the equation
\beq
V^\mu(x)\equiv\omega^{\mu \nu}(x)\partial_\nu H(x) ~~~{\rm or}~~~dH=-i_V\omega
\label{hamvec}\eeq
where the exterior derivative operator $d=\psi^\mu\frac{\partial}{\partial
x^\mu}$ maps $m$-forms to $(m+1)$-forms, and
$i_V=V^\mu(x)\frac{\partial}{\partial\psi^\mu}$
is the interior multiplication operator which contracts differential forms to 1
 lower degree with the vector field $V$. Both $d$ and $i_V$ are nilpotent,
 $d^2=(i_V)^2=0$, and are graded derivations, i.e. they define operators $\cal
 Q$ whose action on differential forms obeys the graded Leibniz rule ${\cal
 Q}(\eta\beta)=({\cal Q}\eta)~\beta+(-1)^m\eta~({\cal Q}\beta)$,
where $\eta$ is an $m$-form. The flows
\beq
\dot x^\mu(t)=V^\mu(x(t))
\label{eqmotion}\eeq
of the Hamiltonian vector field define the classical equations of motion of the
 dynamical system. Note that by definition the symplectic 2-form $\omega$ is
closed, i.e. $d\omega=0$.

We now introduce the Cartan equivariant exterior derivative operator \cite{bv}
\beq
Q_V=d+i_V
\eeq
which is a graded derivation
that maps $m$-forms into the sum of $(m-1)$- and $(m+1)$-forms. If we think of
 commuting, even-degree forms as representing bosons and anti-commuting,
 odd-degree forms as representing fermions, then this suggests that $Q_V$
 represents some sort of supersymmetry operator in the dynamical
 theory. However, unlike the operators $d$ and $i_V$, $Q_V$ is not nilpotent
in general. The square of $Q_V$ is given by the Weil identity
\beq
Q_V^2 \beta =( di_V + i_Vd)\beta = \lie_V \beta
\label{weilid}
\eeq
for the Lie derivative $\lie_V$ along $V$ acting on differential forms. The
form
 $\lie_V\beta$ represents the infinitesimal ($t\to0$) variation as $x(0)\to
x(t)$ of the form $\beta$ under the flows (\ref{eqmotion}) of $V$. The operator
$Q_V$ is therefore nilpotent on
 the kernel $~{\rm ker}~\lie_V=\{\beta\in\Lambda^*\sympl:\lie_V\beta=0\}$ of
the linear derivation $\lie_V$ which represents the subspace
of differential forms which are invariant under the classical dynamics of the
 Hamiltonian system, i.e. for which $\beta(x(t))=\beta(x(0))$. Such forms are
 known as equivariant differential forms \cite{bv}.

{}From the definition (\ref{hamvec}) and the fact that $\omega$ is closed it
 follows that the action in (\ref{clpartpsi}) satisfies
\beq
Q_VS(x,\psi)=(d+i_V)(H+\omega)=dH+i_V\omega\equiv0
\eeq
This means that, if we interpret $Q_V$ as a supersymmetry charge, then the
 action $S$ is supersymmetric and the partition function (\ref{clpartpsi})
 determines a supersymmetric quantum field theory. In this setting, $Q_V$
 determines an $N=\frac{1}{2}$ supersymmetry algebra $Q_V^2=\lie_V$ and the
BRST
 complex of physical states is the space $~{\rm ker}~\lie_V$ of
 equivariant differential forms. In the mathematics literature
 the BRST cohomology of the charge $Q_V$ (i.e. the space of $Q_V$-closed forms
$\eta$, $Q_V\eta=0$, modulo $Q_V$-exact forms $\eta=Q_V\beta$) is called the
$U(1)$-equivariant
 cohomology of $\sympl$ generated by the action of $V$ on $\sympl$
 \cite{ab,bv,blau-thom}. Notice also that, because of the Leibniz rule for
 $Q_V$, the differential form (\ref{alpha}) is also supersymmetric,
 $Q_V\alpha=0$.

We shall now derive a general integration formula for the partition
 function (\ref{clpart}) in terms of geometrical objects on the phase space
 which will allow us to examine its localization features explicitly. Consider
 the integral
\beq
{\cal Z}(s)=\int_\sympl\alpha\e^{-sQ_V\beta}
\label{locint}\eeq
where $\beta$ is an arbitrary globally defined differential form on $\sympl$.
We assume that ${\cal Z}(s)$ is a regular function of $s\in\IR^+$ and
that its $s\to0$ and $s\to\infty$ limits exist. Its $s\to0$ limit is just the
  integral $Z(T)=\int_\sympl\alpha$ of interest while its $s\to\infty$ limit
 represents a localization of (\ref{clpartpsi}) onto the smaller subspace of
 $\sympl$ where $Q_V\beta=0$. Then (\ref{locint}) and the identity
\beq
{\cal Z}(0)=\lim_{s\to\infty}{\cal Z}(s)-\int_0^\infty ds~\frac{d}{ds}{\cal
Z}(s)
\label{zsgen}\eeq
imply that the partition function (\ref{clpartpsi}) can be determined as
\beq
Z(T)=\lim_{s\to\infty}\int_\sympl\alpha
\e^{-sQ_V\beta}+\int_0^\infty ds~\int_\sympl Q_V(\alpha \beta)
\e^{-sQ_V\beta}
\label{halfway}
\eeq
where we have used the fact that $\alpha$ is supersymmetric.

Consider first the last integration in (\ref{halfway}). Since $Q_V$ is a
graded derivation we can integrate by parts to get
\beq
\int_\sympl Q_V(\alpha \beta)
\e^{-sQ_V\beta}= \int_\sympl Q_V \left( \alpha \beta
\e^{-sQ_V\beta} \right) + \int_\sympl \alpha \beta
Q_V \left( \e^{-sQ_V\beta} \right)
\label{2intparts}\eeq
In the first integral on the right-hand side of (\ref{2intparts}) there is an
 $i_V$-exact integrand. Since interior multiplication reduces the order of a
 form by one and integration over the manifold is non-zero only on the
top-degree component of any differential form, it follows that this integral
 vanishes. As for the $d$-exact integration in this same integral, we can use
Stokes' theorem to write it as an integral over the $(2n-1)$-dimensional
boundary $\partial\sympl$ of the manifold $\sympl$. Finally, in the last
integral we recognize the Lie derivative from (\ref{weilid}), and hence
\beq
\int_0^\infty ds~\int_\sympl Q_V(\alpha \beta)
\e^{-sQ_V\beta}=
\int_0^\infty ds~\left\{\oint_{\partial\sympl}\alpha\beta
\e^{-sQ_V\beta}-s\int_\sympl\alpha\beta(\lie_V\beta)
\e^{-sQ_V\beta}\right\}
\label{2ndhalf}\eeq

To carry out the first integration in (\ref{halfway}), we must explicitly
specify the form $\beta$. In principle there are many possibilities for $\beta$
\cite{niemi-tirk,szsem1,blau-thom}, but in
 order to obtain finite results in the limit $s\to\infty$ we need to ensure
that
 the form $Q_V \beta$ has a 0-form component to produce an exponential damping
 factor, since higher order forms will contribute only polynomially due to
 antisymmetry (see (\ref{alpha})). This is guaranteed only if $\beta$ has a
 1-form component. Thus it is only the 1-form part of $\beta$ that will be
 relevant in the following, and so without loss of generality we assume that
 $\beta\equiv B_\mu\psi^\mu$. Furthermore, we need for the 0-form part $V^\mu
B_\mu$ of $Q_V
 \beta$ to attain its global minimum at zero so that the large-$s$ limit of
 (\ref{halfway}) yields a non-zero result. This boundedness requirement is
  equivalent to the condition that the component of $B$ along $V$ has the same
 orientation as $V$. In order to implement such a condition
we need to introduce a globally defined Euclidean-signature metric tensor
 $g_{\mu\nu}(x)$ on the phase space. Then the most general form of $\beta$ up
to
 components orthogonal to $V$ is given by
\beq
\beta (x,\psi)= f(x) g(V,\cdot) \equiv
f(x) g_{\mu \nu}(x) V^\mu(x)\psi^\nu
\label{beta}
\eeq
where $f(x)$ is any strictly-positive smooth-function on $\sympl$.

With this choice for $\beta$ we have $Q_V \beta = K_V + \Omega_V $ where
\beq
K_V = f\cdot g(V,V) \equiv f\cdot g_{\mu \nu}V^\mu V^\nu~~~,~~~\Omega_V
 = d [ f\cdot g(V,\cdot)] \equiv f(2g\cdot\nabla V-\lie_Vg)+(df)g(V,\cdot)
\label{kvomegav}
\eeq
Here $\nabla\equiv d+\Gamma$ is the usual covariant derivative with $\Gamma$
the
 Levi-Civita-Christoffel (affine) connection associated with the Riemannian
metric $g$, and
\beq
(\lie_Vg)_{\mu\nu}=g_{\mu\lambda}\nabla_\nu V^\lambda+g_{\nu\lambda}\nabla_\mu
V^\lambda
\eeq
are the components of the Lie derivative of $g$ along $V$.
We now substitute these identities into (\ref{halfway}) to write the first
 integration there as a sum over the critical point set $I(H)$ which coincides
 with the zero locus of the Hamiltonian vector field $V$,
\beq\new{\begin{array}{lll}
\lim_{s\to\infty}\int_\sympl\alpha \e^{-sQ_V\beta}
&=&\lim_{s\to\infty}\int_{\sympl\otimes T^*\sympl}d^{2n}x~
d^{2n}\psi~\frac{\e^{iT(H+\frac{1}{2}\omega_{\mu\nu}\psi^\mu\psi^\nu)}}
{(iT)^n}\e^{-sf\cdot g_{\mu\nu}V^\mu
V^\nu-\frac{s}{2}(\Omega_V)_{\mu\nu}\psi^\mu\psi^\nu}\\&=&\left(\frac{2\pi
 i}{T}\right)^n\int_{\sympl\otimes T^*\sympl}
d^{2n}x~d^{2n}\psi~\e^{iT(H(x)+\frac{1}{2}\omega_{\mu\nu}(x)
\psi^\mu\psi^\nu)}~f^{-n}(x)~\frac{\delta(V(x))}{\sqrt{\det
g(x)}}\\&&~~~~~~~~~~~~~~~\times~\pfaff~\Omega_V(x)~\delta(\psi)
\\&=&\left(\frac{2\pi i}{T}\right)^n\sum_{p\in I(H)}
\frac{f^{-n}(p) \e^{iTH(p)}}{|\det dV(p)|}~\frac{\pfaff~\Omega_V(p)}{\sqrt{\det
 g(p)}}\end{array}}
\label{zsinf}\eeq

Finally, we can rewrite the determinants in (\ref{zsinf}) using the fact that
 at a critical point $p\in I(H)$
the Hessian of $H$ can be written using the Hamilton equations (\ref{hamvec})
as
\beq
\hess(p)_{\mu\nu}\equiv\partial_\mu\partial_\nu H(p)=-(\partial_\mu
V^\lambda)(p)\omega_{\lambda\nu}(p)
\label{hessatp}\eeq
and likewise from the definition of the 2-form $\Omega_V$ in (\ref{kvomegav})
we have
\beq
g^{\mu\lambda}(p)(\Omega_V)_{\lambda\nu}(p)=f(p)\left\{ 2\omega^{\mu\lambda}(p)
\hess(p)_{\lambda\nu}-g^{\mu\lambda}(p)({\cal L}_Vg)_{\lambda\nu}(p)\right\}
\label{momatp}
\eeq
where $g^{\mu\nu}$ is the matrix inverse of $g_{\mu\nu}$.
Substituting (\ref{hessatp}) and (\ref{momatp}) into the large-$s$
limit integral (\ref{zsinf}) in (\ref{halfway}) and combining this
with (\ref{2ndhalf}) and the choice of $\beta$ in (\ref{beta}), we
 arrive at our final expression for the integral (\ref{clpart}) in
 terms of
geometrical characteristics of the phase space
\beq\new{\begin{array}{lll}
Z(T)&=&\left(\frac{2\pi}{T}\right)^n\sum_{p\in
I(H)}\e^{i\frac{\pi}{4}\eta_H(p)}
\sqrt{\frac{\det\omega(p)}{\det\hess(p)}}\e^{iTH(p)}\sqrt{\det\left(
{\bf1}-\hess^{-1}\omega g^{-1}\lie_Vg/2\right)(p)}\\&&
+\frac{1}{(iT)^n}\int_0^\infty ds~
\oint_{\partial\sympl} \frac{\e^{iTH-sK_V}}{(n-1)!}~ f\cdot g(V,\cdot)
\left(iT\omega-s\Omega_V\right)^{n-1}\\&&-\frac{1}{(iT)^n}\int_0^\infty ds~s
\int_\sympl\frac{\e^{iTH-sK_V}}{(n-1)!}~f^2\cdot g(V,\cdot)(\lie_Vg)(V,\cdot)
\left(iT\omega-s\Omega_V\right)^{n-1}\end{array}}
\label{classpartgen}\eeq
where we have used the fact that the fermion field $g(V,\cdot)$ is nilpotent,
 and the factor $\e^{i\frac{\pi}{4}\eta_H(p)}$ takes into proper account of
the sign of the Pfaffian $\pfaff~dV(p)\sim\sqrt{\det\omega^{-1}\hess(p)}$ in
 (\ref{zsinf}) at each critical point $p\in I(H)$.

The expression (\ref{classpartgen}) represents an alternative to the
conventional loop-expansion \cite{hormander} which explicitly takes into
account the geometric symmetries that make the 1-loop approximation exact.
 It is readily seen that, for closed phases spaces
 ($\partial\sympl=\emptyset$), (\ref{classpartgen}) reduces to
 the Duistermaat-Heckman integration formula (\ref{dhthm}) whenever
 the Hamiltonian vector field $V$ is a conformal Killing vector of a metric
 $g$, i.e.
\beq
\lie_V g = \Lambda g
\label{confkilling}\eeq
which in local coordinates reads
\beq
g_{\mu \lambda} \nabla_\nu V^\lambda +
g_{\nu \lambda} \nabla_\mu V^\lambda =
\frac{1}{n} \left(\nabla_\lambda V^\lambda \right)g_{\mu\nu} =
\frac{1}{n} \left(\nabla_\lambda \omega^{\lambda \rho}
\partial_\rho H \right) g_{\mu \nu}
\label{confkillingloc}\eeq
where the smooth-function $\Lambda(x)=\tr~\nabla V(x)/n$ is fixed by
 contracting both sides of (\ref{confkilling}) with $g^{\mu\nu}$.

Notice that the scaling function $\Lambda(x)$ in (\ref{confkilling})
 vanishes at the critical points of $H$, so that the only possible
global Hamiltonian conformal Killing vectors are those which generate
 global isometries of $g$, $\Lambda(x)\equiv0$ almost everywhere on
$\sympl$, or non-homothetic transformations for which $\Lambda(x)$ is
 a globally-defined non-constant function on $\sympl$. The former case,
 wherein the Hamiltonian vector field is a Killing vector of some
 globally-defined Riemannian metric tensor on $\sympl$, is well-known
 to represent a quite general class of dynamical systems for which
 the Duistermaat-Heckman formula is exact
 \cite{bv},\cite{blau-keski}--\cite{blau-thom}. In that case, we set $f=1$
 in (\ref{beta}) so that then $\lie_V\beta=(\lie_Vg)(V,\cdot)=0$,
 i.e. $\beta\in{\rm ker}~\lie_V$ is a supersymmetric fermion field.
 Then the derivation of (\ref{classpartgen}) also serves to show that the
localization integral (\ref{locint}) coincides with the partition function
(\ref{clpart}) for all $s$. This is because the supersymmetric
action $S-sQ_V\beta$ in ${\cal Z}(s)$ is cohomologous under $Q_V$
 to the supersymmetric action $S$, and the partition function
(\ref{clpartpsi}) depends only on the BRST cohomology class of
$S$, and not on its particular representative. This feature is
often refered to as the equivariant localization principle
 \cite{ab,witten,dlr,szsem1,blau-thom} -- we can topologically
 renormalize the action $S(x,\psi)$ without changing
the value of the integral $Z(T)$ in (\ref{clpartpsi}).
 This can be thought of as a Ward identity associated
with the ``hidden" supersymmetry of the dynamical theory. Notice that if $H$
has no stationary points on the boundary $\partial\sympl$ of $\sympl$, then the
first $s$-integration on the right-hand side of (\ref{2ndhalf}) can be carried
out explicitly and yields
\beq
Z_{\partial\sympl}^{(0)}(T)=\oint_{\partial\sympl}\frac{\e^{iTH}}{g(V,V)}
{}~g(V,\cdot)~\sum_{k=0}^{n-1}\frac{(-1)^k}{(n-k-1)!}\left(\frac{\Omega_V}{K_V}
\right)^k\frac{\omega^{n-k-1}}{(iT)^{k+1}}
\eeq
which represents the additional contribution to the Duistermaat-Heckman
integration formula (\ref{dhthm}) for manifolds with boundary \cite{bv}.

The case of a non-vanishing $\Lambda(x)$ is similar to the isometry case
 from the point of view of the localization mechanism discussed above.
Note that away from the critical points of $H$ we can choose the function
 $f(x)$ in (\ref{beta}) so that $\beta = g(V,\cdot)/g(V,V)$.
With this choice for the localization 1-form $\beta$ it
is easy to show that away from the critical point set of the Hamiltonian
 it satisfies
$\lie_V \beta = 0$. Thus away from the subset $I(H)\subset\sympl$ the
conformal Killing condition can be cast into the same supersymmetric
context as the isometry condition by a rescaling of the metric
 tensor in (\ref{confkilling}),
$g_{\mu\nu}\to G_{\mu\nu}=g_{\mu\nu}/g(V,V)$, for which $\lie_VG=0$.
Of course the rescaled metric $G_{\mu\nu}(x)$ is only defined
 on $\sympl-I(H)$, but all that is needed to establish the localization of
(\ref{clpartpsi}) onto the zeroes of the vector field $V$ (i.e. the
equivariant localization principle) is an invariant metric tensor (or
 equivalently an equivariant differential form $\beta$) which is defined
 everywhere on $\sympl$ except possibly in an arbitrarily small
neighbourhood of $I(H)$ \cite{keski-niemi,blau-thom}.

However, in
contrast to isometry generators the generators of conformal transformations
need not correspond to a global $U(1)$-action on $\sympl$. This is
because although the isometry group of a compact
 space is itself compact, the conformal group need not be.
We might therefore expect that {\it globally} the case of a non-vanishing
 scaling function $\Lambda(x)$ in (\ref{confkilling}) represents a new
sort of localizable Hamiltonian dynamics. To explore this possibility, we now
turn to an
explicit example of a Hamiltonian system which generates non-zero conformal
 motions of a Riemannian metric. We consider the plane
$\sympl=\IR^2$ with its usual flat Euclidean metric which in complex
coordinates is $ds^2=dzd\bar z$. In this case the conformal
Killing equations (\ref{confkillingloc}) become simply $\partial_{\bar z}V^z=
\partial_z V^{\bar z}=0$, and thus the conformal group of the flat plane is
generated by {\it arbitrary} holomorphic vector fields $V^z=F(z),V^{\bar z}=
\bar F(\bar z)$ (these generate the infinite-dimensional classical Virasoro
 algebra). The classical equations of motion determined by the Hamiltonian
 flows of these vector fields are therefore the {\it arbitrary}
analytic coordinate transformations
\beq
\dot z(t)=F(z(t))~~~~~,~~~~~\dot{\bar z}(t)=\bar F(\bar z(t))
\label{confflows}\eeq

We shall now explicitly construct a Hamiltonian system associated with such a
vector field. For definiteness, we consider the conformal Killing vector which
describes
a Hamiltonian system with $n+ 1$ distinct stationary points,
\beq
V^z ~ = ~ i \beta z ( 1 - \alpha_1 z) \cdots ( 1- \alpha_n z)
\label{ckv}
\eeq
at $z=0$ and $z=1/\alpha_i$, where $\beta,\alpha_i\in\IC$. The associated
 scaling function in (\ref{confkilling}) is then
\beq
\Lambda(z,\bar z)=\partial_z V^z+\partial_{\bar z}V^{\bar z}
\label{scalefn}\eeq
The integrability of the Hamiltonian equations (\ref{hamvec}) requires that
the symplectic 2-form be invariant under the flows of $V$, i.e. $\lie_V
\omega=di_V\omega=0$. This leads to the first-order linear partial differential
equation
\beq
V^z\partial_z\omega_{z\bar z}+V^{\bar{z}}\partial_{\bar{z}}\omega_{z\bar z}
=-\Lambda(z,\bar z)\omega_{z\bar z}
\label{heqns}\eeq
where $\omega\equiv\omega_{z\bar z}\psi^z\psi^{\bar z}$. The equation
(\ref{heqns}) is easily solved by separation of variables,
and the solution for the symplectic 2-form with arbitrary separation parameter
$\lambda\in\IR$ is
\beq
\omega_{z\bar z}^{(\lambda)}(z,\bar z)=w_\lambda(z)\bar w_\lambda(\bar
z)/V^zV^{\bar z}
\label{ckvomegax}\eeq
where
\beq
w_\lambda(z)=\e^{i\lambda\int dz/V^z}=\left(\frac{z}{ ( 1 -  \alpha_1 z )^{A_1}
\cdots ( 1 - \alpha_n z)^{A_n}}\right)^{\lambda/\beta}
\label{wlambda}\eeq
and the constants
\beq
A_i( \alpha_1,\dots,\alpha_n) =  \left(\alpha_i\right)^{n-1} ~
\prod_{j\neq i}
 \frac{1}{\alpha_i - \alpha_j}
\eeq
are the coefficients of the partial fraction decomposition
\beq
(V^z)^{-1}=\frac{1}{i\beta}\left(\frac{1}{z}+\sum_{i=1}^n\frac{A_i}{1 -
\alpha_i z}\right)
\eeq

To ensure that (\ref{ckvomegax}) is a single-valued function on $\IC$,
we restrict the $\alpha_k$'s to all have the same phase, so that
$A_i( \alpha_1,\dots,\alpha_n)\in\IR$'s, and the parameter $\beta$ to be
real-valued.
The Hamiltonian equations (\ref{hamvec}) can now be integrated up with the
vector field (\ref{ckv}) and the symplectic 2-form (\ref{ckvomegax}), from
which we find the family of Hamiltonians
\beq
H^{(\lambda)}_{\beta,\alpha_i}(z,\bar z)=\frac{1}{\lambda}\left(\frac{z}{ ( 1 -
 \alpha_1 z )^{A_1} \cdots ( 1 - \alpha_n z)^{A_n}}\right)^{\lambda/\beta}
\left(\frac{\bar z}
{ ( 1 -  \bar{ \alpha}_1 \bar{z} )^{A_1} \cdots ( 1 - \bar{ \alpha}_n
\bar{ z})^{A_n}} \right)^{\lambda/ \beta}
\label{ckvHx}\eeq
To ensure that this Hamiltonian has only non-degenerate
critical points we set $\lambda = \beta$.
 This also guarantees that the level (constant energy) curves of this
Hamiltonian coincide with the curves which are the solutions of the equations
of motion (\ref{confflows}).

 Since the Hamiltonian (\ref{ckvHx}) either vanishes or is infinite on its
critical point set, it is easy to show that the partition function
(\ref{clpart}) is independent of ${\alpha_k}$ and coincides with the
 anticipated result from the Duistermaat-Heckman integration formula
(\ref{dhthm}),
\beq
Z(T) = \int dz~ d\bar{z}~ \omega_{z \bar{z}}^{(\beta)}(z,\bar z) \e^
{i T H_{\beta,\alpha_i}^{(\beta)}(z,\bar{z})}=\frac{ 2 \pi i \beta} { T}
\eeq
This partition function coincides with that of the simple harmonic oscillator
 Hamiltonian $H=\frac{1}{\beta}(p^2+q^2)\sim z\bar z/\beta$. Indeed, if
 we set $\alpha_i=0$, then $\omega^{(\beta)}_{z \bar{z}}$ becomes the
Darboux 2-form and $H^{(\beta)}_{\beta,0}$
the harmonic oscillator Hamiltonian. Furthermore, the scaling
function (\ref{scalefn}) vanishes, the Killing vector $V^z=i\beta z$
generates rotations of the plane, and the Hamiltonian
flows (\ref{confflows}) are the circular orbits $z(t)=\e^{i\beta(t-t_0)}$ about
the origin in the complex
plane of period $2\pi/\beta$. This is the classic example of a dynamical system
with WKB-exact partition function, and moreover it is the unique localizable
system on a homogeneous phase space \cite{szsem1} (i.e. one with
$\nabla\omega=0$, for which the only
possible conformal motions (\ref{confkillingloc}) are isometries).

In fact, we can integrate up the flow equation (\ref{confflows}) in the general
case and we find that the classical trajectories $z(t)$ are determined by the
equation
\beq
 \e ^{ i \beta (t - t_0) } = w_\beta(z(t))=\frac{ z(t)}{
( 1 -  \alpha_1 z(t) )^{A_1 } \cdots ( 1 - \alpha_n z(t))^{A_n} }
\label{conftrans}
\eeq
The coordinate change $z\to w_\beta(z)$ is just the finite conformal
transformation generated by the vector field (\ref{ckv}) and it maps the
dynamical system $(\omega_{z\bar z}^{(\beta)},H_{\beta,\alpha_i}^{(\beta)})$
onto the harmonic oscillator $H\propto w\bar w$, $\omega\propto\psi^w\psi^{\bar
 w}$ with circular classical trajectories $w(t)=\e^{i\beta(t-t_0)}$. This
transformation
is in general multi-valued and has singularities at the critical points
$z=1/\alpha_i$ of the Hamiltonian $H^{(\beta)}_{\beta,\alpha_i}$. It is
therefore not a diffeomorphism of the plane for $\alpha_i\neq0$ and the
Hamiltonian system $(\IR^2,\omega_{z\bar
z}^{(\beta)},H_{\beta,\alpha_i}^{(\beta)})$ is not globally isomorphic to the
simple harmonic oscillator.

The transformations (\ref{conftrans}) are one-to-one in a neighbourhood of
the origin but are one-to-many when the energy of the system is
above a critical value $E_c$.  Asymptotically, the Hamiltonian
$H_{\beta,\alpha_i}^{(\beta)}$ tends to a finite value which is given by
\beq
E_c \equiv \lim_{|z|\to\infty}H_{\beta,\alpha_i}^{(\beta)}(z,\bar z) =
\frac{ 1 }{\beta}\prod_{k=1}^n  ( \alpha_k \bar{\alpha}_k) ^ {- A_k}
\label{limitH}\eeq
If we consider the effect of the change of variable (\ref{conftrans})
on the Hamiltonian system with $H=w_\beta\bar{w}_\beta$, then circular orbits
of energy less than $E_c$ are mapped into closed orbits which are contractible
to the origin and are in a one-to-one correspondence with the domain orbits
in the complex $z$-plane.
As the energy tends towards $E_c$ these orbits grow larger and at the
critical energy they actually reach infinity and return in a finite time
determined by the frequency $\beta$.  Above the critical energy the
transformation (\ref{conftrans}) is in general
one-to-many whereby single orbits
are mapped to distinct orbits about each of the $n$ singular points
$\{z_k=1/\alpha_k\}_{k=1}^n$ of the Hamiltonian. This complicated behaviour of
the conformal flows (\ref{conftrans}) is in marked contrast to the nature of
the harmonic oscillator orbits which always just encircle the origin.

\begin{figure}
\epsfysize= 2.4in
\epsfbox[-200 432 400 720]{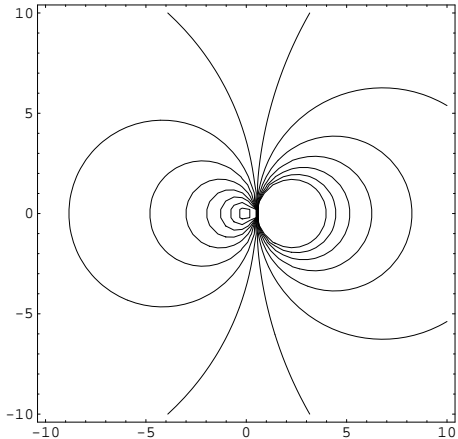}
\caption{Conformal flows for the quadratic generator $V^z=iz(1-z)$.
\label{fig:fig1}}
\end{figure}

These unusual characteristics of the conformal flows (\ref{confflows}) are
best illustrated by  explicit examples.
First, we consider the case when the Hamiltonian vector field is quadratic
in $z$, $V^z = i \beta z ( 1 - \alpha z)$. In this case the flow equation
(\ref{conftrans}) can be solved explicitly to give
\beq
z(t)=\frac{1}{\alpha +\e^{-i\beta(t-t_0)}}
\label{quadrflow}\eeq
The orbit (\ref{quadrflow}) describes a circle in the complex plane
centered, for total energy $\beta H=E$,
at the point $E \bar{\alpha} (E |\alpha|^2 -1 )^{-1}$
and of radius $\sqrt{E} ~ | E |\alpha|^{2} -1 |^{-1}$ (Fig. \ref{fig:fig1}).
In this case the (invertible)
M\"obius transformation (\ref{conftrans}) effectively maps the
 point at $w_\beta=\infty$ to $z=1/\alpha$, and the flows (\ref{quadrflow})
are unbounded and go out to infinity as $E \to 1 /\alpha \bar{\alpha}$. Notice
that this particular example is also applicable to the compactified case where
$\sympl$ is the Riemann sphere $S^2\simeq\IC\cup\{\infty\}$. There
the conformal group is the finite-dimensional Lie group $SL(2,\IC)/\IZ_2\simeq
SO(3,1)$ of
 projective conformal (M\"obius) transformations (i.e. for which $V^z(z)$
 is at most quadratic in $z$) and volume forms contain an additional factor
 $(1+w\bar w)^{-2}$ associated with the compactness of $S^2$. In this case
 the M\"obius
transformation $z\to w_\beta(z)$ above is a diffeomorphism of the Riemann
sphere. As
for the plane, however, the conformal dynamics (\ref{quadrflow}) are quite
different than the isometric dynamics generated by the usual height function
\cite{niemi-pas,keski-niemi,szsem1}, and this is related to the fact that while
 the isometry group $SO(3)$ of $S^2$ is compact, its conformal group is not.
The conformal group structures on spaces like $S^2$ give novel generalizations
of the localizable systems which are usually associated with coadjoint orbits
of the appropriate isometry groups and the quantization of spin systems
\cite{niemi-pas,keski-niemi,szsem1,szsem2}.

\begin{figure}
\epsfysize= 2.4in
\epsfbox[-200 432 400 720]{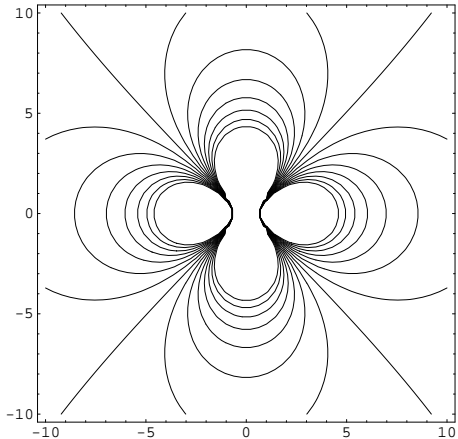}
\caption{Conformal flows for the cubic generator $V^z=iz(1-z^2)$.
\label{fig:fig2}}
\end{figure}

Next, we consider the more complicated example of the
cubic conformal Killing vector $V^z = i \beta z ( 1 - \alpha ^2 z^2)$ for which
the classical trajectories are determined by the equation
\beq
\e^{i\beta(t-t_0)}=\frac{z(t)}{ \sqrt{1-\alpha^2  z^2(t)} }
\eeq
Here the more complicated  trajectories which coincide with the level curves of
the Hamiltonian
exhibit a distinct difference between the usual circular
orbits of the linear (harmonic oscillator) and
quadratic vector field discussed above. Additionally, in this case
for energies above $E_c=1/\beta|\alpha|^2$ we realize a one-to-two mapping
(\ref{conftrans}) of the plane as the point at $w=\infty$ is now mapped to
$z=\pm 1/\alpha$. For the example $\alpha = \beta=1$ depicted in Fig.
\ref{fig:fig2} there is a central
hour-glass shaped region which can be seen to be in a one-to-one correspondence
with the domain orbits. For energies greater than $E_c = 1 $ in this case,
the classical trajectories of the system depend crucially on initial conditions
as there are equivalent orbits about each singularity at $z= \pm 1$.

The apparent
 equivalence between localizable Hamiltonian systems and harmonic oscillator
 Hamiltonians is also observed for those which generate isometries
 \cite{szsem1}. It is a consequence of the fact that these Hamiltonians
generate circle actions, which is the basic ingredient in the
 Duistermaat-Heckman theorem. This large degree of symmetry
in the theory is precisely what is required to reduce the
complicated integrations in (\ref{clpart}) to Gaussian (harmonic
 oscillator) ones and hence render the semi-classical approximation
to the partition function exact. It would be interesting to construct conformal
integrable models on more general
 spaces other than the ones we have considered above where the equivalence with
the harmonic oscillator dynamics was quite transparent because of the
holomorphic-antiholomorphic decomposition of the Hamiltonian system. In the
general case we expect that more complicated Hamiltonians which generate
conformal motions will share common features with those which are isometry
generators, but the dynamics of these systems will be quite different. These
different dynamical structures could play an important role in path integral
localizations which are expressed in terms of trajectories on the phase space
\cite{blau-keski,niemi-tirk}. It would be very interesting to see if these
general conformal symmetries of the classical theory remain unbroken by quantum
corrections in a quantum mechanical path integral generalization. The absence
of such a conformal anomaly could then lead to a generalization of the above
extended localizations to path integral localization formulas. The appearance
of the larger (non-compact) conformal group in certain settings may also lead
to interesting new structures, such as in coadjoint orbit quantization
\cite{niemi-pas,keski-niemi,szsem1,szsem2} or the nonabelian generalizations of
the Duistermaat-Heckman theorem \cite{witten} which employ the full isometry
group of the phase space.

\bigskip
We wish to thank D. Austin, R. Froese, I. Kogan, A. Polyakov, A. Niemi and
O. Tirkkonen for helpful discussions.

\end{document}